\documentstyle[prl,twocolumn,aps,graphicx,floats]{revtex}
\def\bq{\begin{equation}}
\def\eq{\end{equation}}
\def\ba{\begin{eqnarray}}
\def\ea{\end{eqnarray}}

\begin{document}
\thispagestyle{empty}

\newcommand{\sla}[1]{/\!\!\!#1}

\preprint{
\font\fortssbx=cmssbx10 scaled \magstep2
\hbox to \hsize{
\hskip.5in \raise.1in\hbox{\fortssbx Fermilab}
\hfill\vtop{\hbox{\bf }
            \hbox{\today}} }
}

\title{\vspace{-1.0cm}
{\normalsize \rm
\begin{flushright}
Fermilab-Pub-00/082-T \\
\end{flushright} }
A new method for extracting \\
the bottom quark Yukawa coupling \\
at the CERN Large Hadron Collider
}
\author{D.~Rainwater\\[3mm]}
\address{
Theoretical Physics Department, Fermilab, P.O. Box 500, Batavia, IL 60510
}
\maketitle
\begin{abstract}
We propose a new method for measuring the $H\to b\bar{b}$ rate at the CERN 
Large Hadron Collider in a manner which would allow extraction of the $b$ 
quark Yukawa coupling. 
Higgs boson production in purely electroweak $WHjj$ events is calculated. 
The Standard Model signal rate including decays $W\to\ell\nu$ and 
$H\to b\bar{b}$ is 11 fb for $M_H = 120$~GeV. 
It is possible to suppress the principal backgrounds, $W b\bar{b} jj$ and 
$t\bar{t} jj$, to approximately the level of the signal. 
As the top quark Yukawa coupling does not appear in this process, 
it promises a reliable extraction of $g_{Hbb}$ in the context of the 
Standard Model or some extensions, such as the MSSM.
\end{abstract}

\vspace{3mm}


The observation of electroweak symmetry breaking (ESB) in nature with the 
discovery of massive weak bosons is both a triumph and a tribulation of the 
Standard Model (SM) of particle interactions. While the SM has been incredibly 
successful in explaining the data from various experiments, it has also made 
very powerful predictions for other behavior that was later observed. To date 
there are no glaring contradictions with the SM, nor even any moderately 
uncomfortable ones. Despite this success, direct observation of the mechanism 
believed to be responsible for ESB, the Higgs boson, has not been made. 

All quantum numbers of the SM Higgs boson are known from theory except for its 
mass, which is a free parameter. While fits to electroweak (EW) data suggest 
that the Higgs boson is considerably lighter than its theoretical upper limit 
of 1~TeV, probably on the order of 100~GeV~\cite{EW_fits}, the current direct 
search exclusion limit of $M_H \gtrsim 114$~GeV~\cite{LEP_limits} has pushed 
the SM Higgs boson mass above the EW fits central value. One may interpret this 
as a statement that we are on the verge of discovering the Higgs boson in 
present or near-future experiments.

If a new Higgs--like resonance is discovered at either the CERN Large Electron 
Positron collider, Fermilab's Tevatron machine or the soon to be built CERN 
Large Hadron Collider (LHC), it will be up to the LHC to determine the quantum 
numbers of the resonance and state with conviction whether it is the SM Higgs 
boson, a Higgs boson of an extension to the SM, or something else entirely. Of 
primary importance are the spin and couplings, which must be $SU(2)$ gauge 
couplings to the weak bosons and Yukawa couplings to fermions, proportional to 
the mass of the fermion. At the LHC, it will be fairly straightforward to 
determine the spin~\cite{ATLAS} as well as to extract the gauge 
couplings and total width of a SM-like Higgs boson~\cite{ZKNR}. 
Measuring the Yukawa couplings is more difficult. 
Higgs boson decays to fermions are substantial only for $M_H \lesssim 150$~GeV 
(except for $H\to t\bar{t}$, which is significant for $M_H \gtrsim 350$~GeV); 
heavier Higgs bosons decay dominantly to $W,Z$. However, the LHC will be able 
to measure the $\tau$ Yukawa coupling quite easily for 
$M_H \lesssim 150$~GeV~\cite{ZKNR}.
\footnote{Earlier studies proposed using $b\bar{b}h,h\to b\bar{b}$ events to 
to measure $g_{Hbb}$ at the LHC, but only for very large $\tan(\beta)$ in the 
context of the MSSM~\cite{bbh}. However, detector simulations make questionable 
the feasibility of an effective use of this channel in practice~\cite{ATLAS}.}

This is made possible by the cleanliness and ease of observability of Higgs 
boson production in weak boson fusion (WBF), $pp \to qqVV + X \to qqH + X$, 
where the final state quarks appear as large invariant mass, high-pT tagging 
jets at far forward and backward rapidities in the detector, providing a unique 
signature with which to suppress the background rates. This technique would 
allow observation of decays to the final states $\gamma\gamma$ and 
$\tau^+\tau^-$ for $100 < M_H \lesssim 150$~GeV, and $W^+W^-$ for 
$M_H \gtrsim 120$~GeV~\cite{WBF}, with signal to background (S/B) rate ratios 
typically much better than 1/1. This is the only technique in which the decay 
$H\to\tau^+\tau^-$ could be observed in the SM. As such, it is the only method 
with which to extract a Higgs-fermion-fermion coupling. 
The technique of Ref.~\cite{ZKNR}, which we rely on here, involves combining 
information from several Higgs boson channels to measure both the width to 
$W$ pairs, $\Gamma_W$, which contains the $SU(2)$ gauge coupling of the Higgs 
boson to the weak bosons; and the total Higgs boson width, $\Gamma_{tot}$. 
The ratio of the $H\to\tau^+\tau^-$ rate to the $H\to W^+W^-$ rate in WBF is 
then proportional to $\Gamma_\tau / \Gamma_W$, which allows one to determine 
the $\tau$ Yukawa coupling. 

Measurement of additional Higgs Yukawa couplings is highly desirable. For 
example, in the MSSM there are five physical Higgs bosons, two of which have the 
same quantum numbers as the SM Higgs bosons, but can vary in mass, subject to 
the constraint that the lighter of the two states must have a mass 
$M_h \lesssim 135$~GeV. As such, this state will typically have substantial rate 
for decays to fermions. In the MSSM however, Yukawa couplings of up-type and 
down-type fermions can be altered relative to each other already at tree level. 
This characteristic of the MSSM affects also the rare decay modes (e.g. 
$H\to\gamma\gamma$). In addition, large radiative corrections to the Yukawa 
couplings can modify the tree level decay rates considerably~\cite{CMW}, e.g. 
causing ``misalignment'' of the couplings to $b$ quarks and $\tau$ leptons. 
Thus, direct observation of more than one decay mode can provide a constraint 
on the model. Observation of $H\to c\bar{c}$ or other light fermions is not 
likely to be possible at the LHC. $H\to b\bar{b}$ will be extremely difficult, 
but possible in $t\bar{t}H$ associated production for Higgs boson masses below 
$\sim 120$~GeV~\cite{ATLAS}. However, this process suffers from the 
complication that both an up-type and down-type Yukawa coupling are convoluted, 
thus leaving the model largely undetermined.

Measurement of the $H\to b\bar{b}$ rate in WBF, $qqVV\to qqH\to b\bar{b}jj$, 
would be extremely difficult if not impossible, despite the rate being almost an 
order of magnitude larger than for the decay to tau pairs, as it would suffer 
from enormous QCD backgrounds. There is also the issue of triggering on 
$b\bar{b}jj$ events: they do not typically contain a high-$p_T$ lepton, so may 
not pass a $\sum{E_T}$ trigger with great efficiency, thus the events would 
simply not be recorded. (An alternative is to design a dual tagging jet 
trigger.) Thus, another measurement of $H\to b\bar{b}$ involving a measured 
production coupling would be highly desirable.

We seek a process that provides a high-$p_T$ lepton in addition to the far 
forward/backward tagging jets and the Higgs boson. The ideal choice is $WHjj$ 
production. Four classes of Feynman diagrams contribute to $pp\to WHjj + X$: 
WBF $H$ production with $W$ bremsstrahlung off a quark leg; WBF $H$ production 
with additional $W$ emission off the $t$-channel weak boson pair; and WBF $W$ 
production and $W$ bremsstrahlung where the Higgs boson is radiated off the $W$. 
Note that the $WHjj$ events we consider are {\it not} QCD corrections to $WH$ 
associated production, but are pure EW processes; QCD corrections to $WH$ events 
will ultimately constitute an enhancement of the signal, but will typically not 
survive the tagging jet cuts or minijet veto and so are neglected here, a 
conservative approximation.

We calculate the cross section for $pp\to WHjj + X$ at the LHC, 
$\sqrt{s} = 14$~TeV, using full tree level matrix elements for all EW 
subprocesses, including finite width and off-shell effects for $W\to\ell\nu$ 
($\ell = e,\mu$), and finite width effects for $H\to b\bar{b}$. The matrix 
elements were generated by {\sc madgraph}~\cite{Madgraph}. The Higgs boson NLO 
decay and total widths are corrected via input from {\sc hdecay}~\cite{Hdecay}. 
CTEQ4L structure functions~\cite{CTEQ} are employed with a choice of 
factorization scale $\mu_{f_i} = p_{T_i}$ of the outgoing tagging jets. 
To provide realistic resolution of the $b\bar{b}$ invariant mass, gaussian 
smearing of final state particle four-momenta is employed according to ATLAS 
expectations~\cite{ATLAS}. We do not decay the bottom quarks explicitly, but do 
include a parameterized energy loss distribution to make a more realistic 
simulation of observed final state momenta, overall missing momentum and 
$b\bar{b}$ invariant mass. As some Feynman diagrams with a t-channel photon 
contribute, the total cross section, shown as a function of $M_H$ in line 1 of 
Table~\ref{sigrate}, is calculated with an explicit initial-final state quark 
pair $Q^2_{ij} > 100$~GeV$^2$ to avoid the singularity from the photon 
propagator. This cut introduces a small uncertainty for the total rate without 
cuts, $\approx\pm 15\%$ for varying the Q$^2$ cut by a factor of 2 (1/2). 
It does not, however, affect the cross sections with cuts.

\begin{table}
\caption{Cross sections (fb) for the $WHjj$ signal as a function of Higgs boson 
mass. Shown on successive lines are the inclusive rate without $W$ or $H$ 
decays, the Higgs boson branching ratio to $b\bar{b}$, and the inclusive rate 
multiplied by the $H\to b\bar{b}$ and $W\to e\nu,\mu\nu$ branching ratios.}
\vspace{0.1in}
\label{sigrate}
\begin{tabular}{l|ccccc}
$M_H$                     & 110  & 120  & 130  & 140  & 150  \\
\hline
inclusive                 &  84  &  80  &  76  &  72  &  70  \\
$B_H(b\bar{b})$           & 0.77 & 0.67 & 0.52 & 0.33 & 0.17 \\
$B_W\cdot B_H\cdot\sigma$ & 13.9 & 11.2 &  8.6 &  5.1 &  2.5 \\
\end{tabular}
\vspace{-0.15in}
\end{table}

The total signal rate appears to be large enough to obtain a significant data 
sample. However, to determine whether this measurement is realistic, we 
calculate the cross section for the main background which can mimic the signal. 
The largest resonant backgrounds are QCD and EW $WZjj;Z\to b\bar{b}$ production, 
but in these cases the $Z$ pole is well-separated from the Higgs boson resonance 
so the overlap should be minimal given the superior detector jet resolutions. 
Thus, we ignore these backgrounds for the present viability check and instead 
concentrate on the largest backgrounds to this signal: nonresonant QCD 
$W b\bar{b}jj$ production, and $t\bar{t}+jets$ events, where both $W$'s from the 
top quarks decay leptonically ($e$ or $\mu$) and one of the leptons is too low 
in $p_T$ to be observed; we take this cut to be $p_T(l,min) < 10$~GeV for the 
simple check here. The latter events consist of QCD corrections to $t\bar{t}$ 
production, but are completely perturbative in the phase space region of 
interest, as the QCD radiation can appear in the detector as far 
forward/backward tagging jets. In addition, there are tree level processes that 
do not correspond to initial or final state gluon radiation. Other potential 
backgrounds are primarily irreducible, or fake signatures, which are naturally 
expected to be subdominant to continuum production.

We calculate the $W b\bar{b}jj$ and $t\bar{t}jj$ rates using exact tree-level 
matrix elements, constructed using {\sc madgraph}~\cite{Madgraph} for the 
former, and the latter from Ref.~\cite{Stange}. We include top quark and $W$ 
leptonic decays to $e,\mu$ in the matrix elements. 
CTEQ4L structure functions~\cite{CTEQ} are employed throughout. We take the 
factorization scale for the $W b\bar{b}jj$ background the same as the signal, 
and the renormalization scale $\mu_r = \sqrt{\mu_{f,1}\cdot\mu_{f,2}}$. For 
the $t\bar{t}jj$ background, $\mu_f = min(E_T)$ of the jets/top quarks, and 
renormalization scale $\mu_r = E_T(jet/top)$, with one factor of $\alpha_s$ 
taken from each of the outgoing jets/top quarks. In all cases, 
$\alpha_s(M_Z) = 0.118$ with 1-loop running.

The basic WBF signature requires the two tagging jets to be at high rapidity 
and in opposite hemispheres of the detector, and the $H,W$ decay products to be 
central and in between the tagging jets. The kinematic requirements for the 
``rapidity gap'' level of cuts are as follows:
\ba
\label{eq:gap}
& p_{T_j} \geq 30~{\rm GeV} \, , \; \; |\eta_j| \leq 5.0 \, , \; \;
\triangle R_{jj} \geq 0.6 \, , \nonumber \\
& p_{T_b} \geq 15~{\rm GeV} \, , \; \; |\eta_b| \leq 2.5 \, , \; \;
\triangle R_{jb} \geq 0.6 \, , \nonumber \\
& p_{T_\ell} \geq 20~{\rm GeV} \, ,\; \;
|\eta_{\ell}| \leq 2.5 \, , \; \; 
\triangle R_{j\ell,b\ell} \geq 0.6 \, , \nonumber \\
& \eta_{j,min} + 0.7 < \eta_{b,\ell} < \eta_{j,max} - 0.7 \, , \nonumber \\
& \eta_{j_1} \cdot \eta_{j_2} < 0, \; \; 
\triangle \eta_{tags} = |\eta_{j_1}-\eta_{j_2}| \geq 4.4 \, .
\ea
The results for $M_H = 120$~GeV are shown in the first column of 
Table~\ref{cuts}.
At this level the backgrounds are already somewhat manageable, but we observe 
that the QCD $W b\bar{b}jj$ background is dominated by low invariant masses of 
the tagging jet pair and low-$p_T$ $b$ jets, so we impose a minimum tagging 
dijet mass and an additional staggered $p_T(b)$ cut to reduce this contribution:
\bq
\label{eq:mjjptb}
m_{jj} > 600 \, {\rm GeV} \; , \quad
{p}_T(b_1,b_2) > 50,20 \, {\rm GeV} \; .
\eq
The $m_{jj}$ cut is also somewhat effective against $t\bar{t}jj$ events, as 
shown in the second column of Table~\ref{cuts}. Furthermore, there are two 
strikingly different characteristics of the signal v. the $t\bar{t}jj$ 
background: the latter events have significantly higher $\sla{p}_T$ on average; 
and they do not exhibit a Jacobian peak in the $m_T(\ell,\sla{p}_T)$ 
distribution, a characteristic of $W$ decays. Both features are due to the fact 
that by suppressing observation of the second charged lepton, the neutrino from 
that $W$'s decay has significantly enhanced transverse momentum, which is 
unobserved, and greatly distorts the $m_T(\ell,\sla{p}_T)$ distribution. 
We choose maximum cutoff values for both observables as follows:
\bq
\label{eq:ptmt}
\sla{p}_T < 100 \, {\rm GeV} \; , \; \; \; 
m_T(\ell,\sla{p}_T) < 100 \, {\rm GeV} \; .
\eq
The result is shown in the third column of Table~\ref{cuts}.

\begin{table*}[htb]
\caption{Cross sections (fb) for the $WHjj$ signal with $M_H = 120$~GeV, and 
QCD $W b\bar{b}jj$ and $t\bar{t}jj$ principal backgrounds, at various levels 
of cuts. Also shown are the progression of S/B, statistical significance in 
Gaussian sigma, and an estimate of uncertainty in the measured cross section. 
100 fb$^{-1}$ of data for each of two experiments is assumed, as in 
Ref.~\protect\cite{ZKNR}. A mass bin $100 < m_{b\bar{b}} < 130$~GeV is 
implicit for all levels of cuts, which captures about $80\%$ of the signal. 
The systematic uncertainty is taken to be $20\%$.}
\vspace{0.15in}
\label{cuts}
\begin{tabular}{l|ccccc}
cuts level & Eq. (\ref{eq:gap}) 
           & + Eq. (\ref{eq:mjjptb}) 
           & + Eq. (\ref{eq:ptmt}) 
           & + veto 
           & $\times \epsilon_{\sc ID}$ = 0.25 \\
\hline
$WHjj$ signal              & 1.4  & 1.4 & 1.1 & 0.81 & 0.20   \\
$Wb\bar{b}jj$ bkg          & 8.8  & 5.7 & 4.3 & 1.30 & 0.32   \\
$t\bar{t}jj$ bkg           & 4.9  & 3.7 & 1.2 & 0.35 & 0.09   \\
\hline
S/B                        & 1/10 & 1/7 & 1/5 & 1/2  & 1/2    \\
$\sigma_{Gauss}$           &      &     &     &      & 4.4    \\
$\triangle\sigma / \sigma$ &      &     &     &      & 35$\%$ \\
\end{tabular}
\vspace{-0.1in}
\end{table*}

A final cut that may be utilized is to reject any candidate event if it contains 
additional central QCD (jet) activity of moderate $p_T$: a 
minijet~\cite{MJorig}. Studies of the minijet rate for WBF, EW and QCD events 
can be found in Refs.~\cite{minijet,WBF} and references therein. Here, we simply 
apply the results from those studies. The probability of a signal event 
surviving a minijet veto, $p_T^{veto}(j) > 20$~GeV, is estimated to be $75\%$, 
slightly lower than that typical of WBF Higgs boson events, because the $W$ 
bremsstrahlung components of $WHjj$ events can slightly enhance the minijet 
activity. The probability of a $t\bar{t}jj$ background event surviving a minijet 
veto is much lower, only about $30\%$; previous studies have indicated it may be 
even better than this for $t\bar{t}+jets$ events, but we choose to remain 
conservative for our proof of concept estimates here. This allows us to achieve 
a S/B rate of 1/2 for $M_H = 120$~GeV, considering only the two major 
backgrounds for this demonstration.

At this stage, shown in the fourth column of Table~\ref{cuts}, the situation 
appears quite good. However, in reality only about $25\%$ of these events will 
be captured in the data sample due to detector efficiencies. We take the 
expected values for ATLAS and CMS to be $86\%$ for each tagging jet, $95\%$ for 
the charged lepton, and $60\%$ for each $b$ quark tag. We assume 100~fb$^{-1}$ 
of data for each of two experiments, as in Ref.~\cite{ZKNR}. Our result at this 
point is comparable to that expected for the $t\bar{t}H; H\to b\bar{b}$ channel 
with $M_H = 120$~GeV: ATLAS expects the significance to be $3.6\sigma$, with 
S/B $\sim 1/3$~\cite{ATLAS}. However, the $b\bar{b}$ mass peak in $t\bar{t}H$ 
events does not clearly stand out, and again, the value $g_{Hbb}$ is not easily 
extractable due to convolution with the top quark coupling.

\begin{figure}[tb]
\includegraphics[width=5.6cm,angle=90]{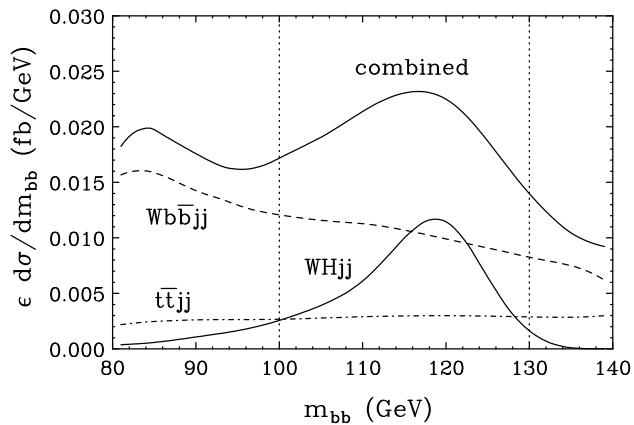} \\
\caption{$b\bar{b}$ invariant mass distribution of the $WHjj$ signal 
($M_H = 120$~GeV), $W b\bar{b}jj$ and $t\bar{t}jj$ backgrounds after all cuts, 
a minijet veto and efficiencies as discussed in the text. The combined signal 
and backgrounds are also shown. Vertical dotted lines denote the mass bin used 
for calculating statistical significance of the signal.}
\label{fig:mbb}
\end{figure}

To illustrate that our method would not be simply a counting experiment, rather 
a distinct resonance could be observed, Fig.~\ref{fig:mbb} shows the invariant 
mass spectrum for the tagged $b$ quark pair. The $W b\bar{b}jj$ and $t\bar{t}jj$ 
background distributions combined are monotonically decreasing above 
$m_{b\bar{b}} = 80$~GeV, allowing the signal to present a clear peak in the 
spectrum for an intermediate mass Higgs boson.

We can make a preliminary rough estimate of the uncertainty in the measurement 
by calculating 
$\triangle\sigma /\sigma \approx \sqrt{N_S + N_B}/N_S \oplus 20\%$, including a 
$20\%$ systematic uncertainty in the backgrounds. This yields about a $35\%$ 
overall uncertainty in the cross section measurement for $M_H = 120$~GeV. As 
this error will dominate the extraction of $\Gamma (H\to b\bar{b})$, we may 
expect the overall measurement error for the partial width to $b$ quarks to be 
${\cal O} (50\%)$ for a Higgs boson of this mass. More precise estimates must 
wait for a more complete consideration of the signal and 
backgrounds~\cite{WHjj}.

We have demonstrated the feasibility of a measurement of the $H\to b\bar{b}$ 
decay rate at the LHC in a moderate background environment with reasonable 
luminosity, in a way that would allow extraction of the bottom quark Yukawa 
coupling. The technique shows promise for Higgs boson masses in the range of 
applicability to the MSSM, but would also be accessible in the SM or other 
SM-like extensions. More importantly, it does this independently of up-type 
Yukawa couplings (e.g. top quark), thus reducing model dependence. Due to the 
accuracy with which $g_{HWW}$, $g_{H\tau\tau}$ and the total Higgs boson width 
will be measured at the LHC~\cite{ZKNR}, $g_{Hbb}$ can be determined by taking 
either the ratio $\Gamma_b/\Gamma_W$ or $\Gamma_b/\Gamma_\tau$, depending on 
the Higgs boson mass and how SM-like the observed state is. 
To be sure, this channel will not provide a precision measurement, but all that 
is needed to constrain models other than the SM, or to rule out regions of the 
MSSM parameter space, is a nonzero width measurement. A more detailed study of 
this process including other backgrounds is underway~\cite{WHjj}. Additional 
statistical significance may be added by using $\gamma Hjj$ production, which 
is also being studied. Early indications are that the data sample would 
approximately double using both production modes.


\acknowledgements
We thank U.~Baur, S.~Dawson and D.~Zeppenfeld for constructive comments and 
U.~Baur, S.~Parke and T.~Plehn for a critical review of the manuscript.
Fermilab is operated by URA under DOE contract No.~DE-AC02-76CH03000.


\end{document}